\documentclass{article}

\usepackage{graphicx}

\author{Hal Finkel}
\title{Stochastic Evolution of Graphs using Local Moves}

\begin{document}
\maketitle

\begin{abstract}
Inspired by theories such as Loop Quantum Gravity, a class of stochastic graph dynamics was studied in an attempt to gain a better understanding of discrete relational systems under the influence of local dynamics. Unlabeled graphs in a variety of initial configurations were evolved using local rules, similar to Pachner moves, until they reached a size of tens of thousands of vertices. The effect of using different combinations of local moves was studied and a clear relationship can be discerned between the proportions used and the properties of the evolved graphs. Interestingly, simulations suggest that a number of relevant properties possess asymptotic stability with respect to the size of the evolved graphs.
\end{abstract}

\section{Introduction}
The dynamics of quantum-gravitational theories such as Loop Quantum Gravity~\cite{smolin04} can be formulated as local moves operating on a special class of labeled graphs. It would be interesting to know whether or not a similar set of local moves operating on generic graphs will generate an ensemble of graphs similar to that assumed by the such quantum-gravitational theories. To that end, unlabeled graphs in a variety of initial configurations were evolved using generalized Pachner moves. Theories of space-time generally assume underlying graphs which are, in an appropriate sense, approximately manifold-like. As it turns out, however, stochastic application of local moves does not seem to generate graphs which are approximately manifold-like. Although the graphs produced do not seem directly interpretable as manifold-like structures, simulations suggest that the ensemble of such graphs is distinguishable from the ensemble of random graphs with the same size and average degree. Certain properties of the graphs produced by the evolution clearly depend on the proportions of different local moves used to generate the evolution. Additionally, simulations suggest that these same properties possess asymptotic stability with respect to the size of the evolved graphs.

\section{Definitions}
A graph is defined to be a pair $G = (V, E)$, where V is a set of vertices, and E is a set of vertex pairs called edges. The degree of a vertex is the number of edges incident on that vertex. The graphs described in this paper are unlabeled and contain no self-loops (edges connecting a vertex to itself) or duplicate edges (edges connecting the same two vertices as some other edge).

\subsection{Generalized Pachner Moves}
\begin{figure}[ht]
\caption{Expansion (``1-to-d'') Pachner Move. $d = 3$ is shown here.}
\label{fig:p1t3}
\begin{center}
\includegraphics[scale=0.7]{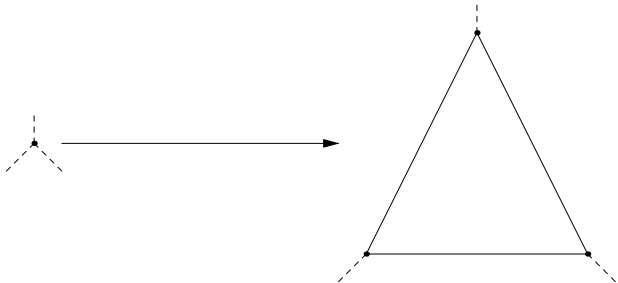}
\end{center}
\end{figure}

\begin{figure}[ht]
\caption{Contraction (``d-to-1'') Pachner Move. $d = 3$ is shown here.}
\label{fig:p3t1}
\begin{center}
\includegraphics[scale=0.7]{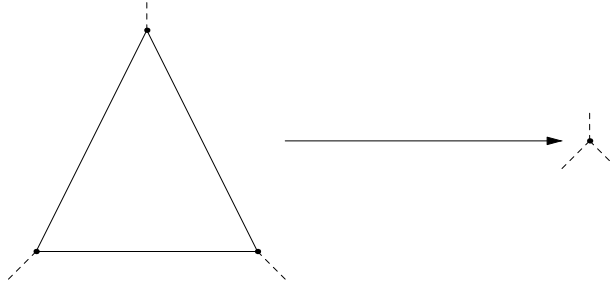}
\end{center}
\end{figure}

\begin{figure}[ht]
\caption{Exchange Pachner Move}
\label{fig:pexch}
\begin{center}
\includegraphics[scale=0.5]{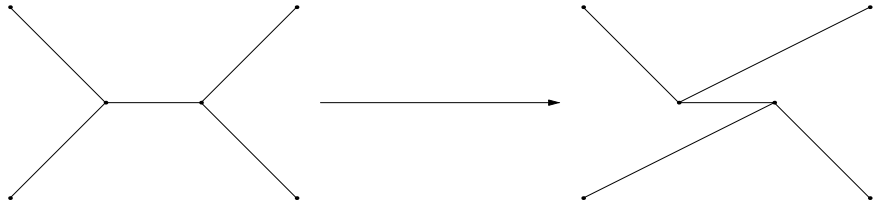}
\end{center}
\end{figure}

\begin{figure}[ht]
\caption{Another Exchange Move}
\label{fig:pexch2}
\begin{center}
\includegraphics[scale=0.5]{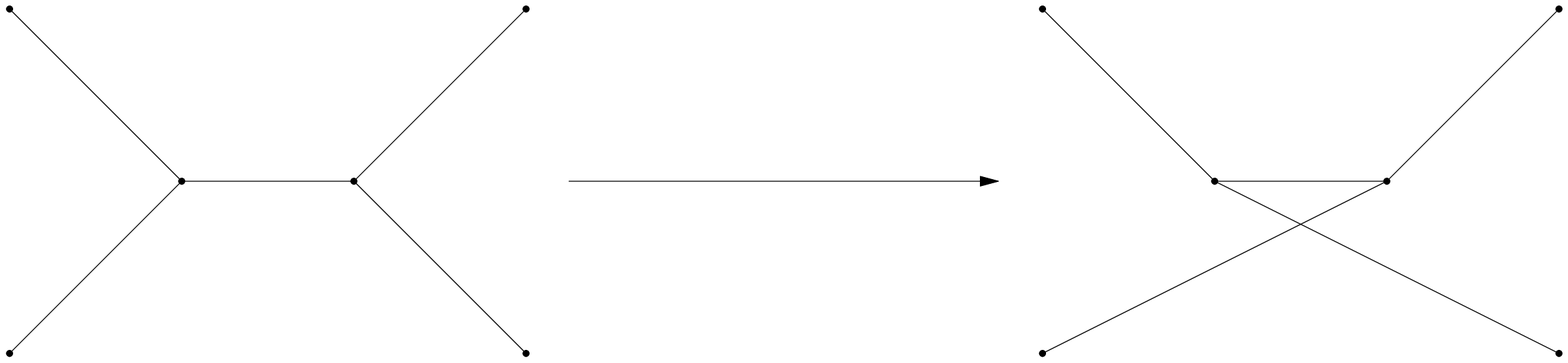}
\end{center}
\end{figure}

For planar trivalent graphs, some combination of the three Pachner Moves can convert any graph into any other graph. The graph dynamics discussed here is a combination of three distinct transformations which are each generalizations of one of the three Pachner Moves. As with traditional Pachner moves, one is an expansion operation, one a contraction operation, and one an exchange operation. The expansion move involves replacing a vertex with a d-simplex (A d-simplex is a set of d vertices where each vertex is connected to all of the other $d-1$ vertices, i.e. a triangle for $d = 3$) (Figure~\ref{fig:p1t3}). The contraction move is the inverse of the expansion move (Figure~\ref{fig:p3t1}). The exchange move exchanges the endpoints of edges both adjacent to a third edge (Figure~\ref{fig:pexch}). Specifically, given some edge $E_1 = (V_1, V_2)$ and edges $E_2 = (V_1, V_3)$ and $E_3 = (V_2, V_4)$ replace $E_2$ and $E_3$ with $E_2' = (V_1, V_4)$ and $E_3' = (V_2, V_3)$. Note that the exchange moves do not preserve planarity or any manifold-like structure (Figure~\ref{fig:pexch2}).

When applying an expansion move, vertices with a degree less than $d$ were not considered for replacement. The old incoming edges were distributed among the new vertices such that each new vertex in the inserted d-simplex was assigned at least one of the old edges.

\subsection{Dimension}
The statistical dimension of the produced graphs was computed. Let $d(v, w)$ be the number of edges in the shortest path between vertex v and vertex w. Let $V(v, r)$ be the number of vertices w such that $d(v, w) \leq r$. If, for a generic vertex v, $V(v, r) = Cr^\alpha$ for some $\alpha$ then $\alpha$ is the statistical dimension. It should be noted that, for an arbitrary graph, the statistical dimension, as specified here, is not well defined.

\subsubsection{Analytic Computation}
It is possible to compute the function $V(v, r)$ for some generic vertex v in a connected random graph with n vertices and m edges.

Let $\bar{d}$ be the average degree of a vertex. Then:

\begin{equation}
	\bar{d} = \frac{2m}{n}
\end{equation}

The vertices in the graph can be divided into equivalence classes based on the length of the shortest path to vertex v. For any given path length $r$ the size of the equivalence class will be denoted $A(r)$. It follows immediately that:

\begin{equation}
	V(r) = V(r-1) + A(r)
\end{equation}

and also that the number of vertices not within a distance $r-1$ of v is:

\begin{equation}
	R(r) = n - V(r-1)
\end{equation}

$s(r)$ is the average number of ``free edges'' per vertex at a distance of $r$.

\begin{equation}
	s(r) = f(r-1) / A(r-1)
\end{equation}

$A(r)$ is the number of vertices at a distance of $r$.

\begin{equation}
	A(r) = R(r) (1 - (1 - \frac{s(r)}{R(r)})^{A(r-1)})
\end{equation}

$f(r)$ indicates the number of ``free edges'', meaning the number of edges connecting vertices at a distance of r to vertices at a distance of $r+1$.

\begin{equation}
	f(r) = (\bar{d} A(r) - f(r-1))(1 - \frac{1}{2}(\frac{A(r-1) - 1}{R(r) - 1}))
\end{equation}

\begin{figure}[ht]
\caption{Plot (Using Analytic Formula) of $\ln(V(r))$ vs. $\ln(r)$ for Random Connected Graphs}
\label{fig:r4}
\begin{center}
\includegraphics[scale=0.5]{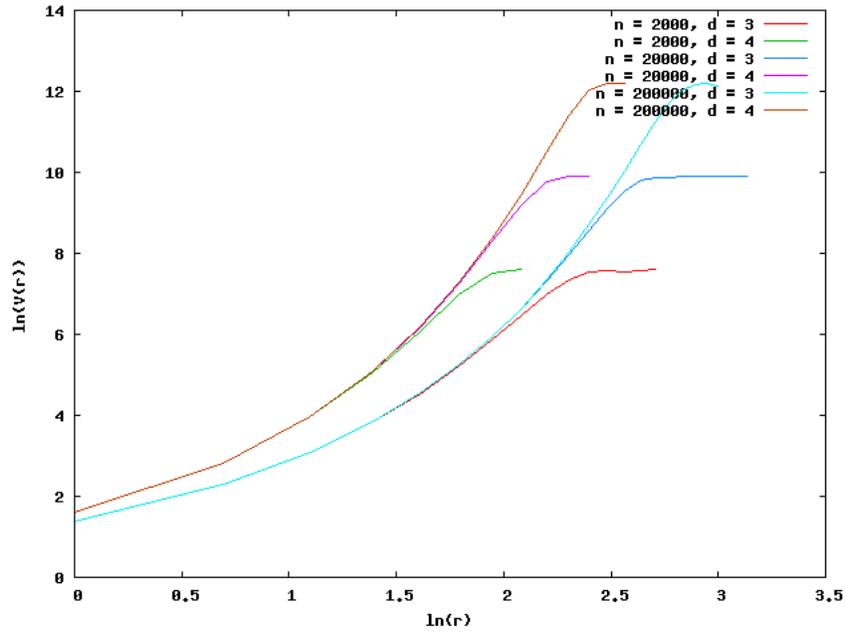}
\end{center}
\end{figure}

Plots of this relation for various values of $n$ and $d$ are shown in Figure~\ref{fig:r4}. It is interesting to note that a generic feature of the random connected graphs is a small statistical dimension and small distances and a larger dimension at larger distances.

\subsection{Non-local Edges}
An edge is said to be ``non-local'' if it is not part of any short cycle (where short means the length of the cycle is less than or equal to some $r_l$). Intuitively, this should mean the edge connects locations in the graph which are ``far away''. If the number of non-local edges is small compared to the total number of edges, then it may be possible to view the graph as some ``local'' structure (such as a regular lattice) combined with some additional distribution of edges. 

\section{Description}
\subsection{Input Parameters}
The simulations used to investigate the properties of the chosen graph dynamics depended on only a few input parameters. These parameters were:
\begin{itemize}
\item $d$: The number of vertices in the simplex added during an expansion move.
\item $r_l$: The length of the largest cycle to consider local.
\item $R$: The ratio of exchange moves to expansion moves.
\item $R_c$: The ratio of contraction moves to expansion moves.
\end{itemize}

\subsection{Graph Pictures Coloring}
The graph pictures which appear in this paper were drawn using an algorithm which treats each edge as a spring with some spring constant and solves for the equilibrium configuration.

The edges are colored as follows (where $r_l$ = 5):
\begin{itemize}
\item Red - Initial edges which are ``local''
\item Magenta - Initial edges which are ``non-local''
\item Black - Edges added by expansion moves which are ``local''
\item Green - Edges added by expansion moves which are ``non-local''
\end{itemize}

\subsection{Initial Conditions}
\begin{figure}[ht]
\caption{Random Connected Graph ($n = 200$ and $m = 400$)}
\label{fig:nl6-00}
\begin{center}
\includegraphics[scale=0.5]{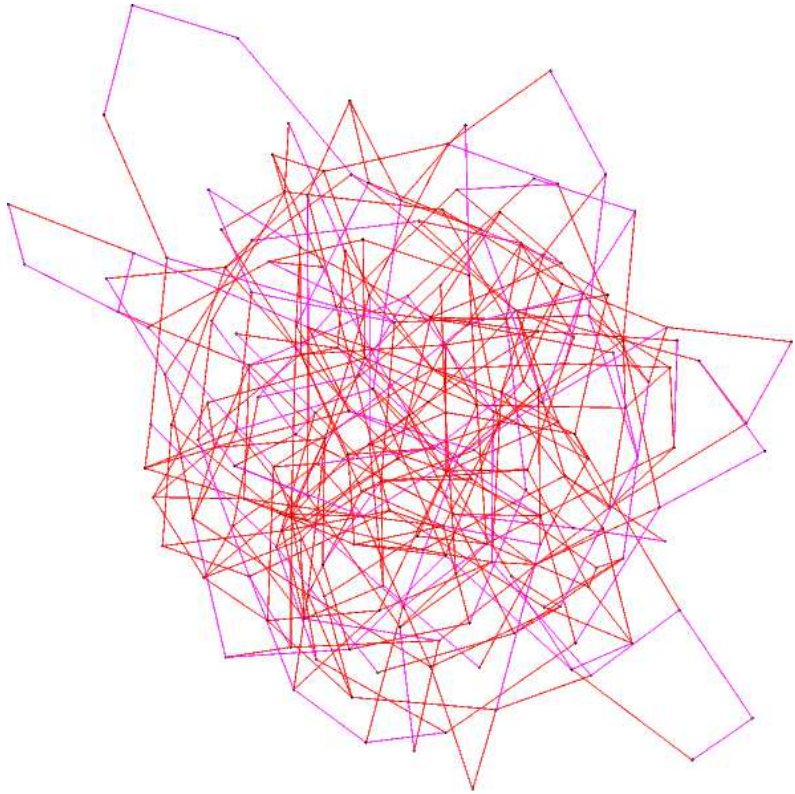}
\end{center}
\end{figure}

Many different starting configurations were tested including a single triangle, regular lattices, almost complete graphs ($\bar{d} = (1 - \epsilon)*(n - 1)$) and random connected graphs. The exact starting configuration seemed to have very little long-term influence on the simulation results. An example of a random connected graph used as an initial configuration is pictured in Figure~\ref{fig:nl6-00}.

\subsection{Simulation Outputs}
\begin{figure}[ht]
\caption{Evolved Random Graph (Evolved using $R = 0$ to $n = 2200$ and $m = 3400$)}
\label{fig:nl6-05}
\begin{center}
\includegraphics[scale=0.5]{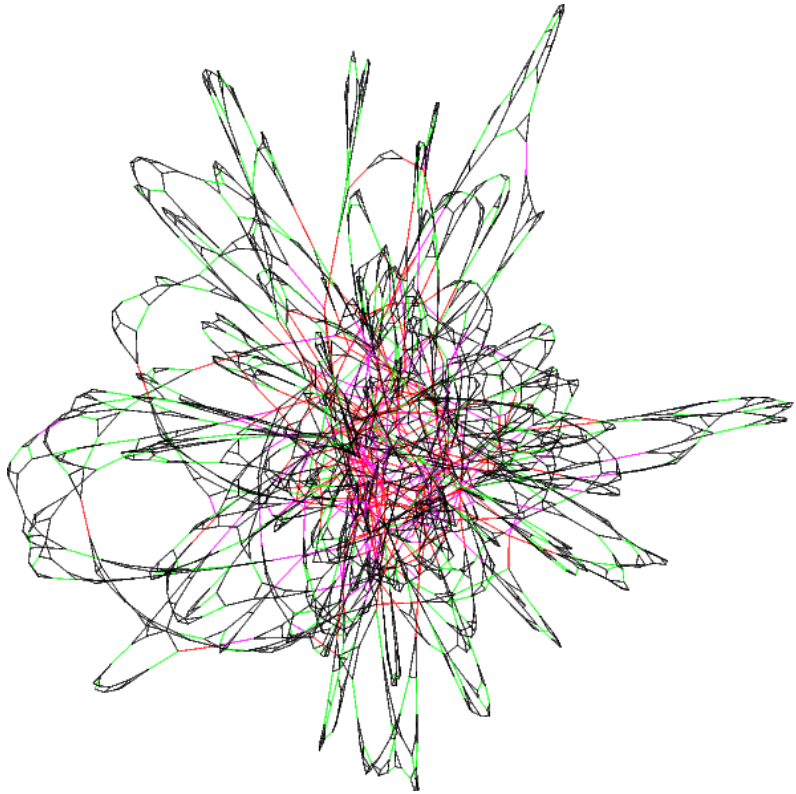}
\end{center}
\end{figure}

\begin{figure}[ht]
\caption{Center Region of Evolved Random Graph (Evolved using $R = 0$ to $n = 2200$ and $m = 3400$)}
\label{fig:nl6-05-2}
\begin{center}
\includegraphics[scale=0.5]{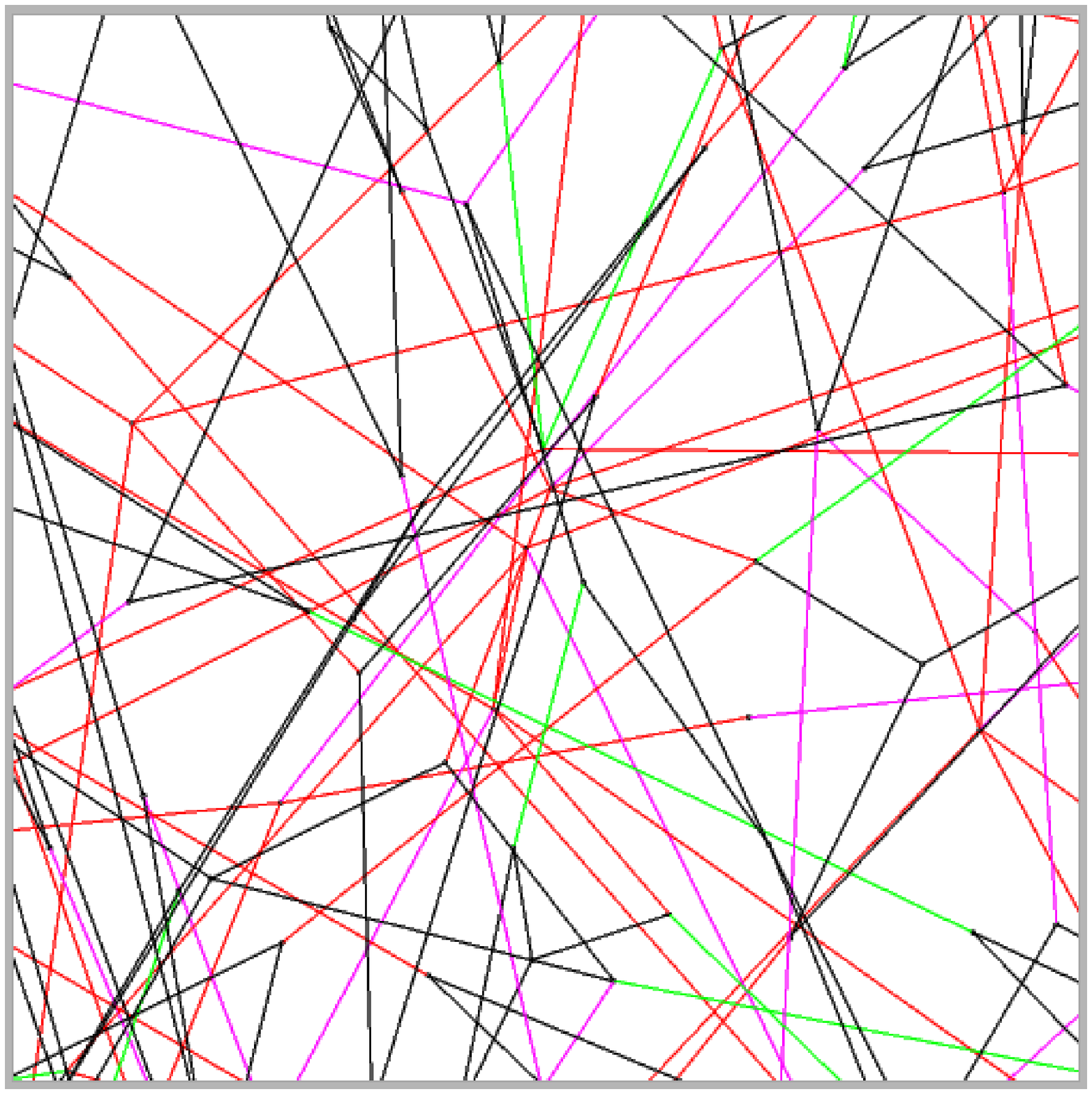}
\end{center}
\end{figure}

\begin{figure}[ht]
\caption{Spike of Evolved Random Graph (Evolved using $R = 0$ to $n = 2200$ and $m = 3400$)}
\label{fig:nl6-05-1}
\begin{center}
\includegraphics[scale=0.5]{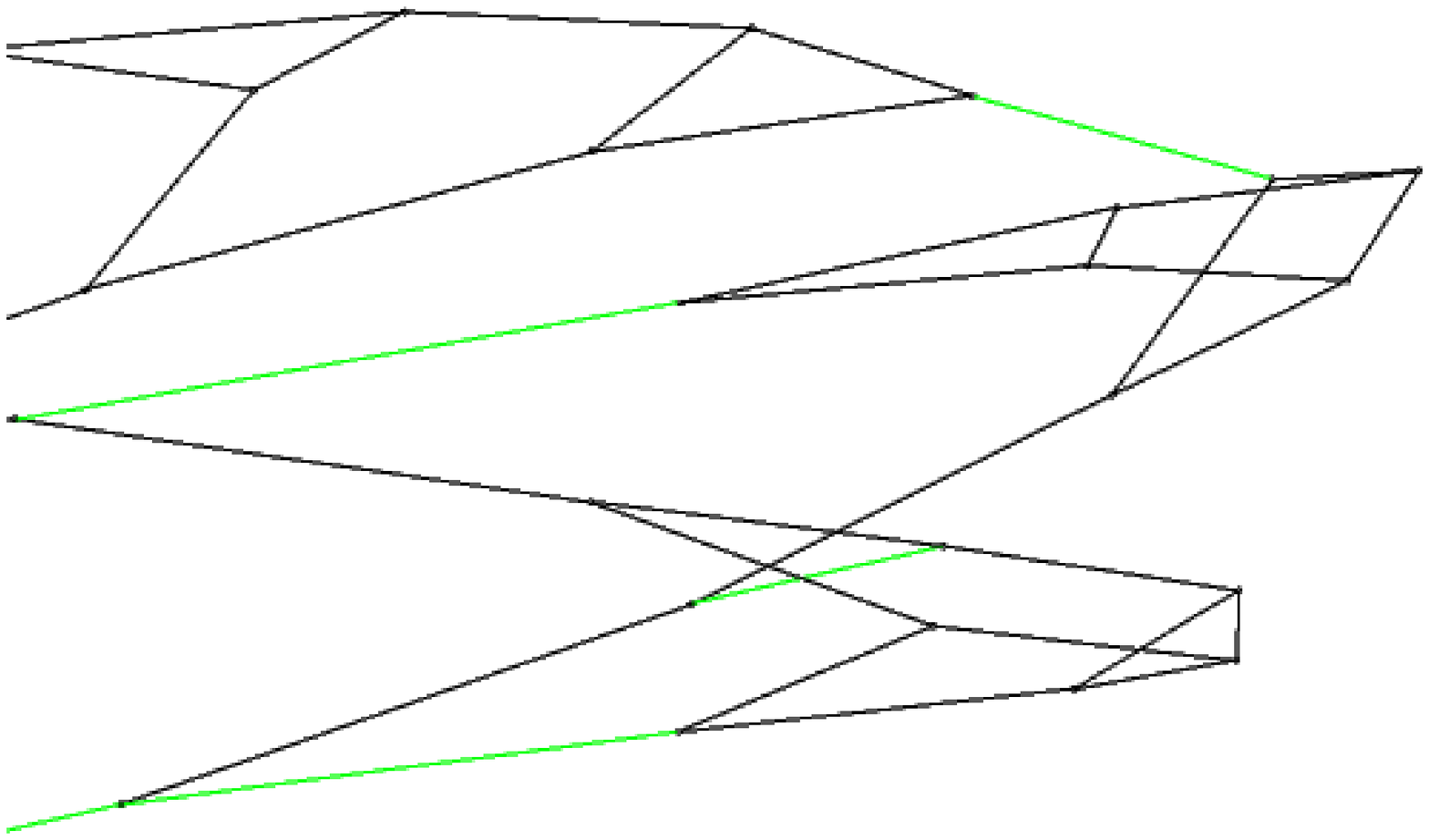}
\end{center}
\end{figure}

\begin{figure}[ht]
\caption{Evolved Random Graph (Evolved using $R = 100$ to $n = 1800$ and $m = 2800$)}
\label{fig:nl3-04}
\begin{center}
\includegraphics[scale=0.5]{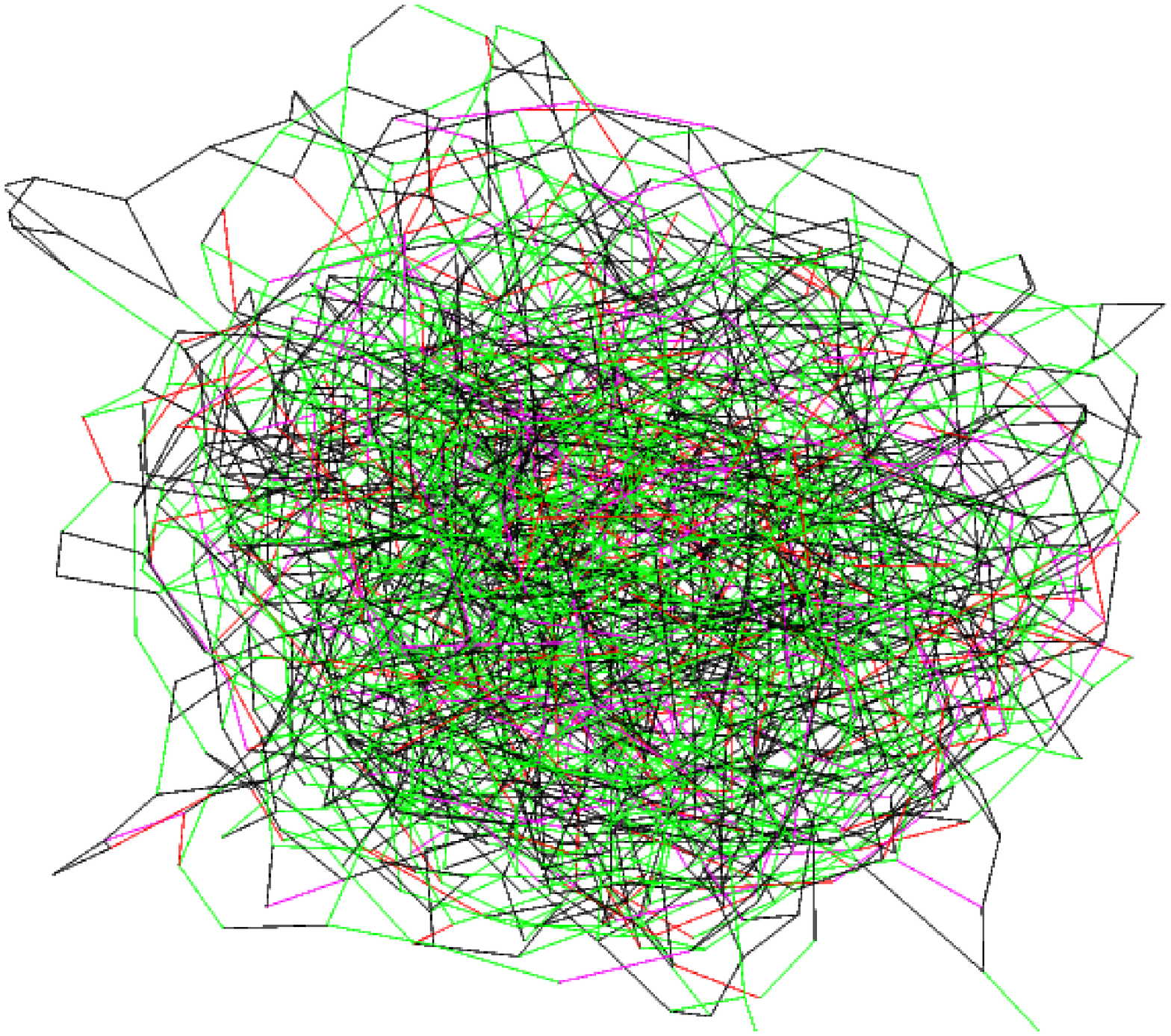}
\end{center}
\end{figure}

\begin{figure}[ht]
\caption{Central Region of Evolved Random Graph (Evolved using $R = 100$ to $n = 1800$ and $m = 2800$)}
\label{fig:nl3-04-1}
\begin{center}
\includegraphics[scale=0.5]{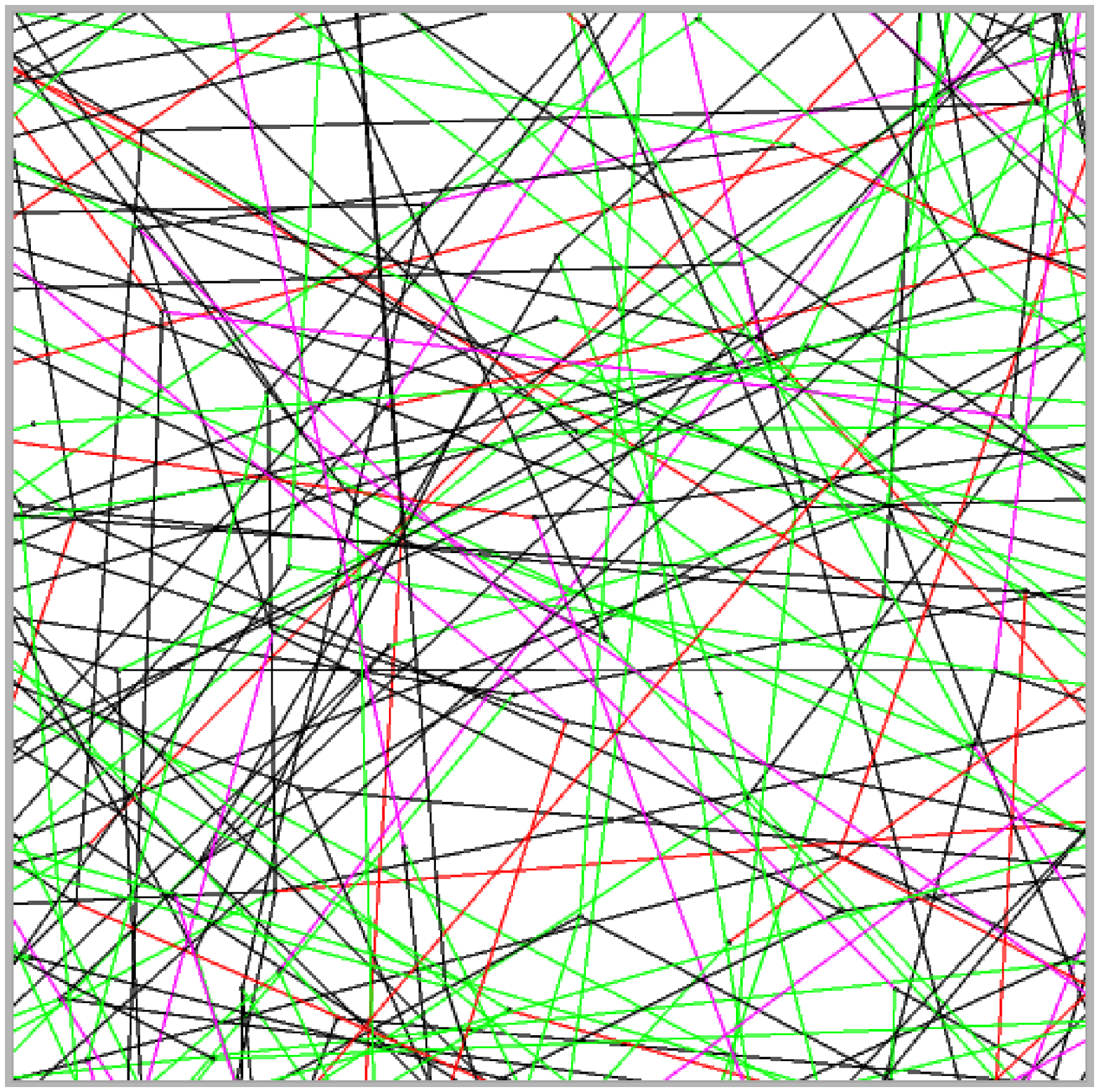}
\end{center}
\end{figure}

For the purpose of this study, the graphs were evolved until they contained tens of thousands of vertices (except for cases when $R_c = 1$). Unfortunately, due to the slow convergence of the graph-drawing algorithm, graphs of that size could not be rendered in a practical manner. However, the properties of the graphs being evolved often converged well before reaching several thousand vertices, and graphs of that size could be feasibly rendered. Figure~\ref{fig:nl6-05} shows the result of the graph pictured in Figure~\ref{fig:nl6-00} evolved using $R = 0$ to where $n = 2200$ and $m = 3400$. Figure~\ref{fig:nl6-05-2} shows a portion of the dense center region and Figure~\ref{fig:nl6-05-1} shows the end of one of the spikes. Approximately $14.7\%$ of the edges are ``non-local.'' Figure~\ref{fig:nl3-04} shows the result of the graph pictured in Figure~\ref{fig:nl6-00} evolved using $R = 100$ to where $n = 1800$ and $m = 2800$. Figure~\ref{fig:nl3-04-1} shows a portion of a central region. Approximately $42.5\%$ of the edges are ``non-local.'' Notice that the addition of the exchange moves seems to suppress the formation of the spiky structures.

\section{Results}
\subsection{Dimension}
\begin{figure}[ht]
\caption{$\ln(V(r))$ vs. $\ln(r)$ for graph evolved with $R = 1$}
\label{fig:dim.out.2.avg.single}
\begin{center}
\includegraphics[scale=0.5]{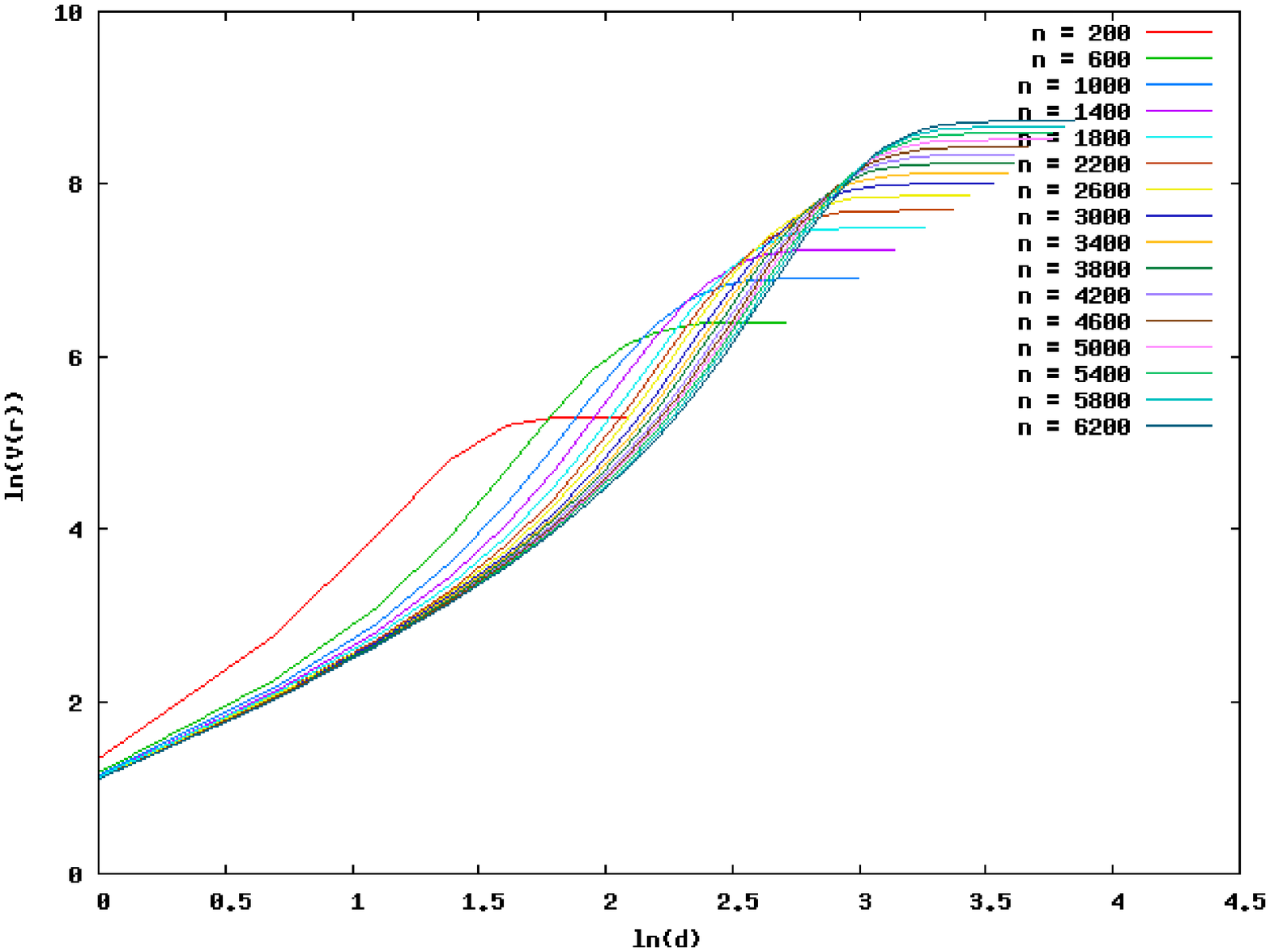}
\end{center}
\end{figure}

\begin{figure}[ht]
\caption{$\ln(V(r))$ vs. $\ln(r)$ for graph evolved with $R = 100$}
\label{fig:dim.out.3.avg.single}
\begin{center}
\includegraphics[scale=0.5]{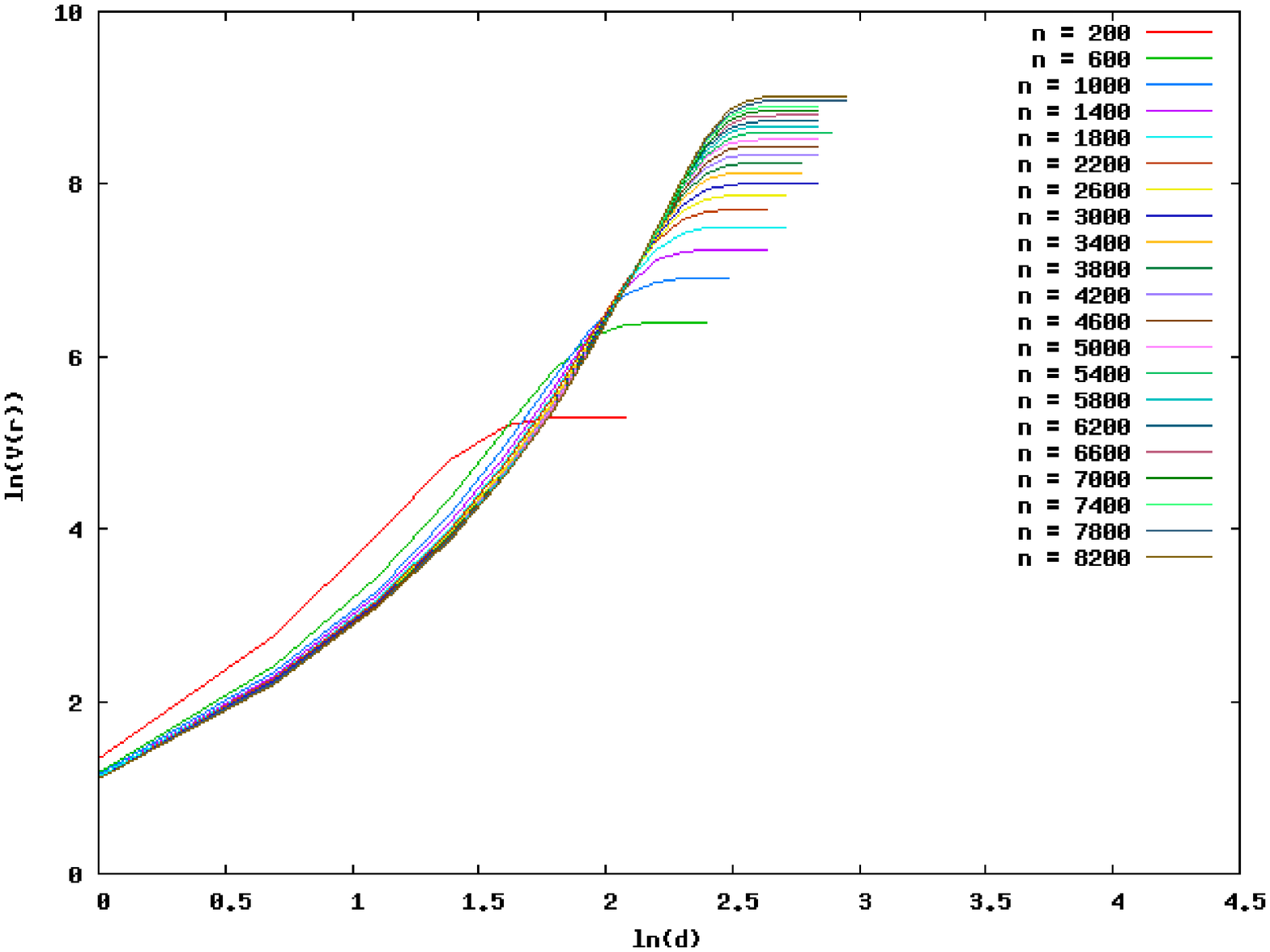}
\end{center}
\end{figure}

The graphs which resulted from the evolution bore little resemblance to regular lattices or any other structure with a statistical dimension constant with scale. $V(r)$ for these graphs is qualitatively similar to that for a random connected graph as can be seen in Figure~\ref{fig:dim.out.2.avg.single} and Figure~\ref{fig:dim.out.3.avg.single}, which show $\ln(V(r))$ vs. $\ln(r)$ for an evolving graph with $R = 1$ and $R = 100$ respectively (a regular lattice of dimension $d$ would appear as a line of slope $d$ in those figures).

\begin{figure}[ht]
\caption{Small-scale dimension vs. R}
\label{fig:ex_v_dim}
\begin{center}
\includegraphics[scale=0.5]{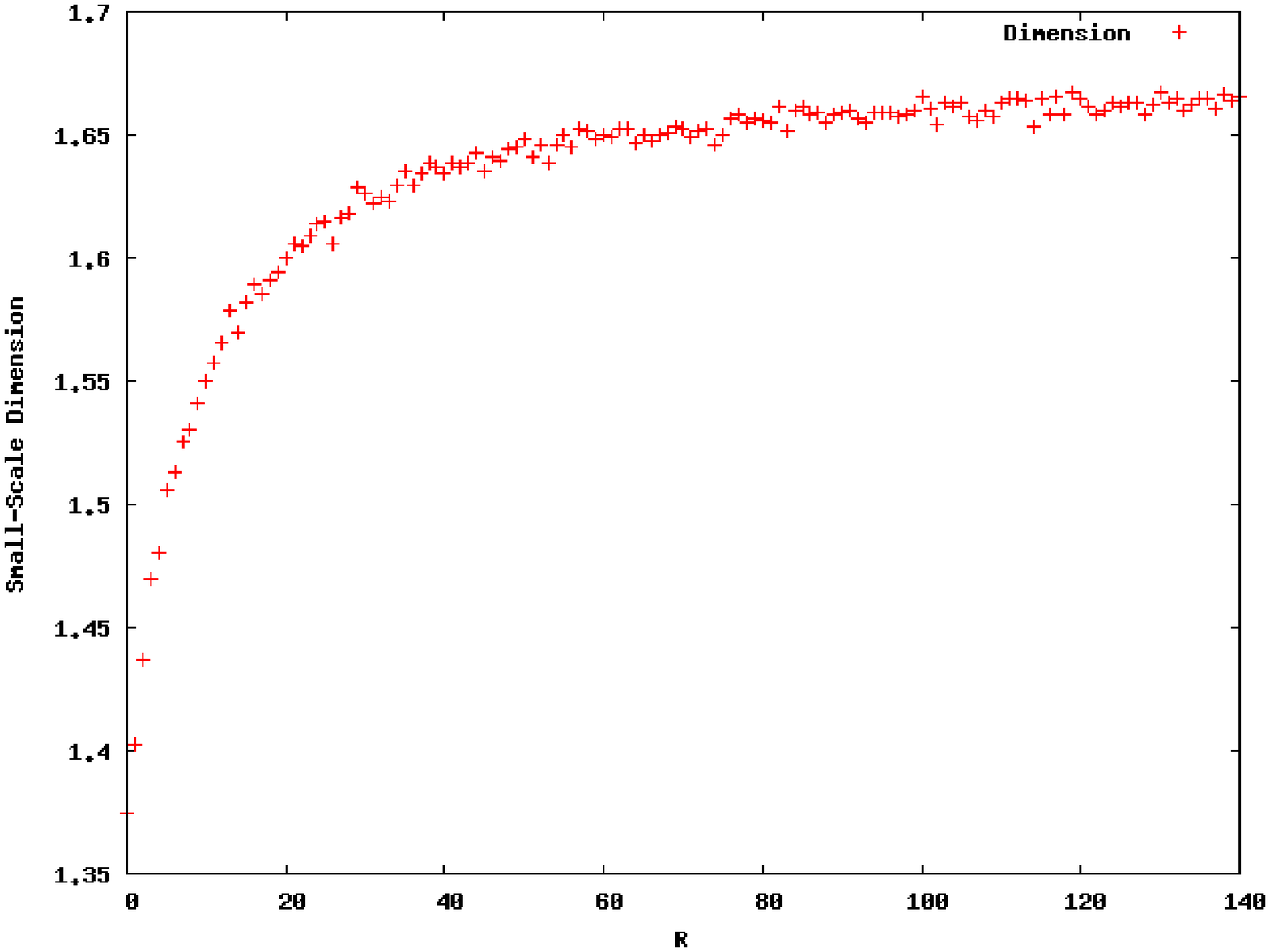}
\end{center}
\end{figure}

Noting that the dimension on small scales is closer to $1$ in Figure~\ref{fig:dim.out.2.avg.single} where $R = 1$ compared to Figure~\ref{fig:dim.out.3.avg.single} where $R = 100$. This can be understood by noting that the graphs evolved with $R = 1$ were dominated by spiky structures which are nearly one-dimensional, whereas evolution with larger values of $R$ suppressed the spike formation. Figure~\ref{fig:ex_v_dim} shows the approximate slope of $\ln(V(r))$ for small $r$ (as determined by a heuristic algorithm) vs. $R$. After some value of $R$ the dynamics appears to become ``exchange dominated,'' whereby further increasing $R$ does not significantly change the structure of the graph. This can also be seen using other measures (such as graph diameter).

\subsection{Non-Local Edges}
\begin{figure}[ht]
\caption{The fraction of non-local edges vs. $n$ ($r_l = 5$)}
\label{fig:f9_fnle}
\begin{center}
\includegraphics[scale=0.5]{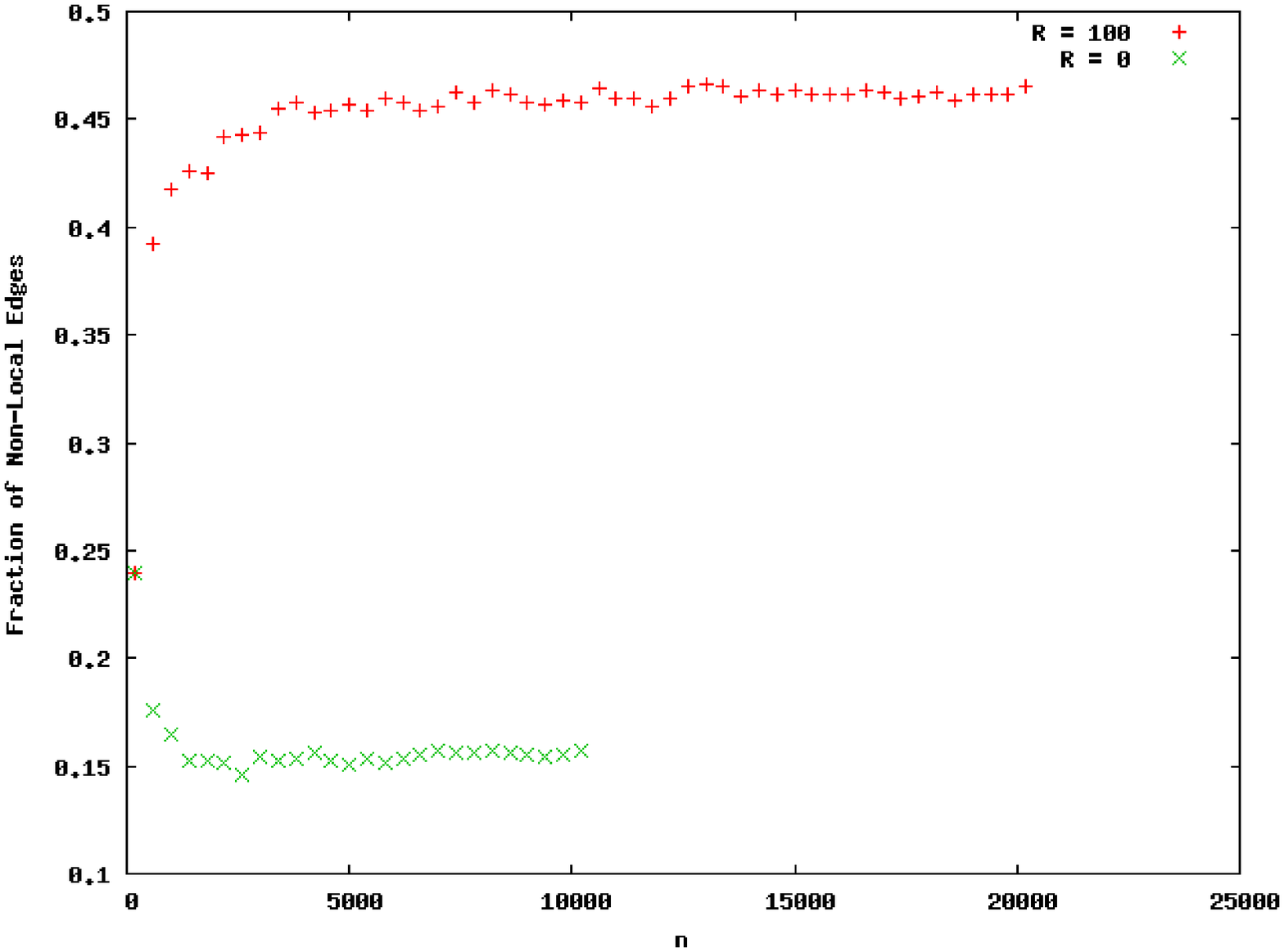}
\end{center}
\end{figure}

\begin{figure}[ht]
\caption{The fraction of non-local edges vs. $r_l$ ($n = 5400$)}
\label{fig:nld_v_nl}
\begin{center}
\includegraphics[scale=0.5]{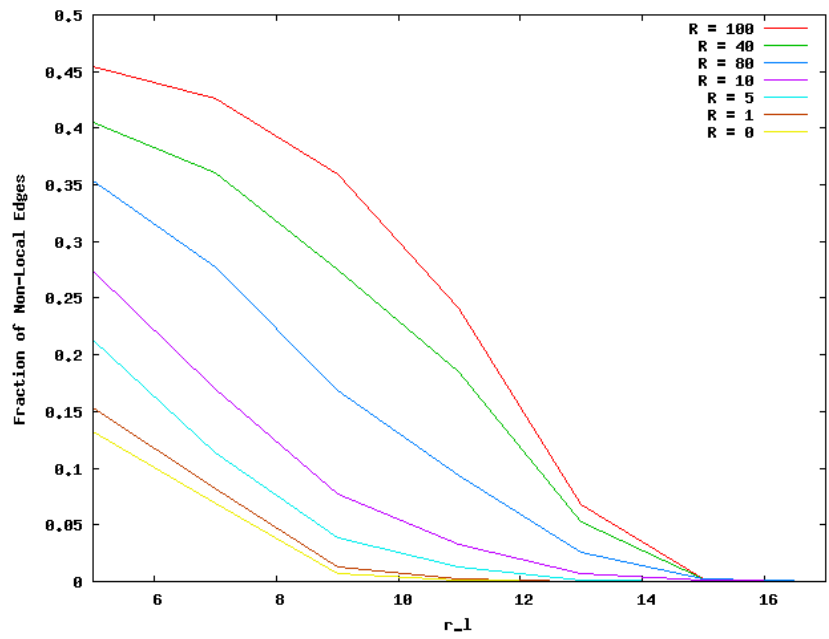}
\end{center}
\end{figure}

\begin{figure}[ht]
\caption{The fraction of non-local edges vs. $r_l$ ($R = 100$)}
\label{fig:nld_v_nl_100}
\begin{center}
\includegraphics[scale=0.5]{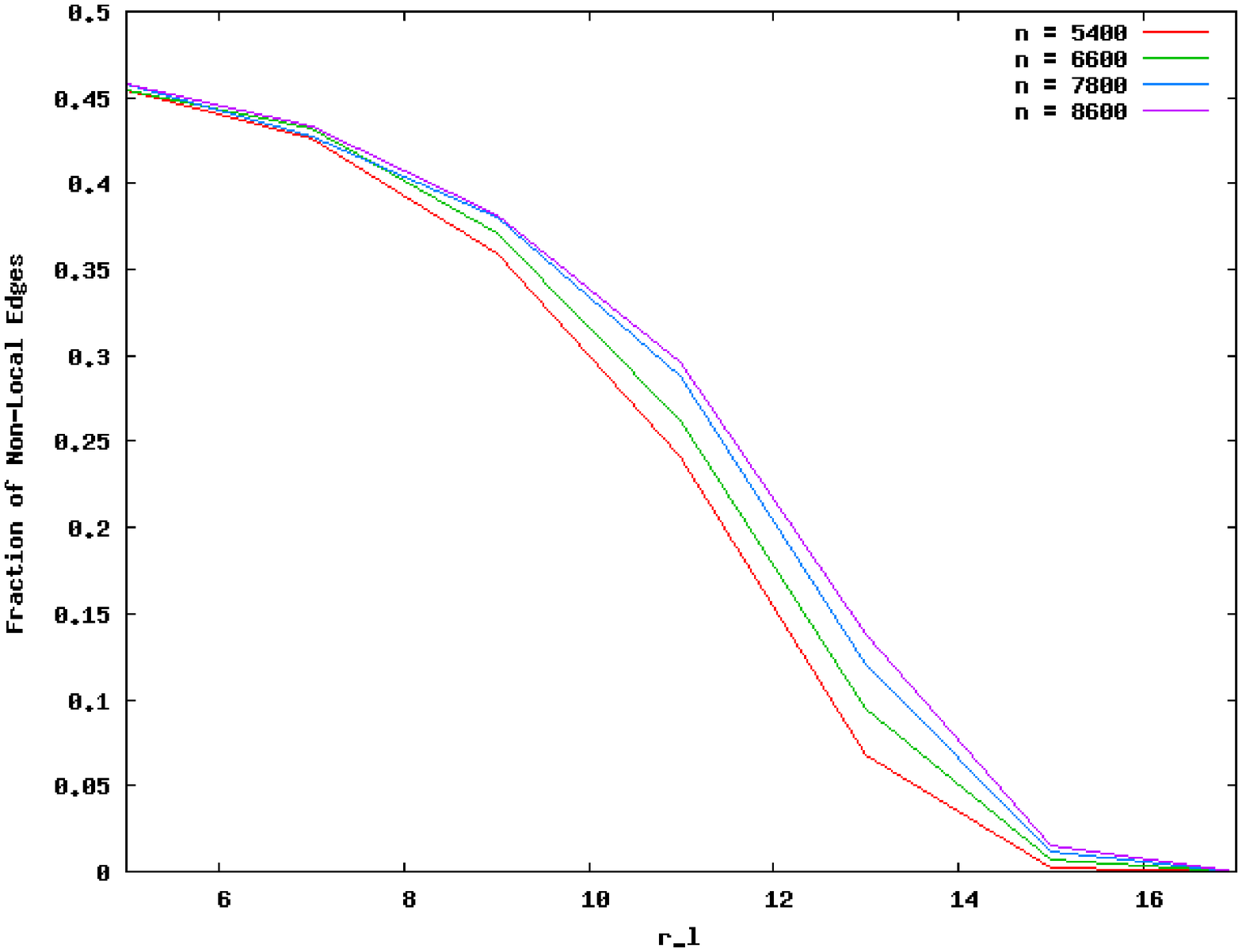}
\end{center}
\end{figure}

Statistical dimension and similar measures do not completely characterize the structure of a graph. The percentage of edges which are ``non-local'' can provide an indication of how much the graph resembles a ``local'' structure such as a regular lattice. Figure~\ref{fig:f9_fnle} shows a plot of the fraction of non-local edges vs. $n$ for some graph evolution (for $r_l = 5$). For any value of $R$ it is possible to identify the asymptotic value of the fraction of non-local edges. Figure~\ref{fig:nld_v_nl} shows the fraction of non-local edges vs. $r_l$ for some fixed graph size. It should be noted that the fraction has only reached its asymptotic value for $r_l$ less than the graph diameter. Figure~\ref{fig:nld_v_nl_100} shows the fraction of non-local edges vs. $r_l$ for a graph evolving using $R = 100$.

\section{Conclusion}
The properties of unlabeled graph evolution under the action of local moves have been investigated and several interesting properties seem to be asymptotically stable with respect to the number of vertices of the graph. Evolving the graphs with $0 < R_c < 1$ did not significantly affect the long-term results (larger values of $R_c$, as should be expected, result in graphs of static or decreasing size). Also, evolving the graphs with $d > 3$ results in qualitatively similar behavior (although, for example, the asymptotic fraction of non-local edges changes as a function of $d$).


\section{Acknowledgments}
I would like to thank Lee Smolin for his constant helpful advice and input into many aspects of this project. I would also like to thank Fotini Markopoulou-Kalamara, Maya Paczuski and Peter Grassberger for their, often critical, commentary and assistance. Finally, I would like to thank the Perimeter Institute for Theoretical Physics for granting me office space, resources and an intellectually-stimulating working environment.


\begin{thebibliography}{77}
\small

\bibitem{smolin04} Smolin, Lee.
{\it An invitation to loop quantum gravity,}
hep-th/0408048.

\end{thebibliography}
\end{document}